\documentclass[%
twocolumn,
amsmath, amssymb,
aps
]{revtex4}

\usepackage{bm}
\usepackage{amssymb}	
\usepackage{amsthm}	
\usepackage{mathrsfs}
\usepackage[flushleft]{threeparttable}
\usepackage{amsmath}
\usepackage{tikz,scalerel}
\newcommand{\beq}{\begin{equation}}
\newcommand{\eeq}{\end{equation}}
\newcommand{\bea}{\begin{eqnarray}}
\newcommand{\eea}{\end{eqnarray}}
\newcommand{\ur}{\mathrm{u}}

\begin{document}

\title{Hidden Parameters in  Heisenberg's and Landau-Peierls 
Uncertainty Relations,\\ Speed of Virtual Particles  and
Space-Time Metric Types}%

 \author{B.~B.~Levchenko}%
\email{levtchen@mail.desy.de }

\affiliation{%
D.V. Skobeltsyn Institute of Nuclear Physics, M.V. Lomonosov Moscow State University,\\ 
 119991 Moscow, Russian Federation\\
		}%

\begin{abstract}
In quantum field theory, virtual particles are carriers of relativistic wave fields, participate in vacuum fluctuations, and are a consequence of the implementation of the quantum uncertainty mechanism.
Virtual particles, playing a fundamental role in our understanding and description of interaction processes, are themselves inaccessible to direct observation and measurement of their dynamic characteristics.
We prove that the well-known Heisenberg uncertainty relations 
and Landau-Peierls uncertainty relations implicitly contain  new ``hidden'' angular variables. 
On the basis of the relations obtained, we derive a formula  for estimating the 
speed $u$ of a virtual particle in indirect measurements. This formula is applicable in cases where the final state is characterized by an incomplete set of variables.  
Moreover, we present arguments that the speed of a virtual particle 
 is a sensitive indicator of  non-linearity  of the process dynamics 
and  propose a  method for revealing the type of space-time geometry 
in the interaction region of high-energy particles. 

\end{abstract}



\maketitle

\section{Introduction}
%
Many physical processes in quantum systems can be correctly described 
only by introducing different types  of intermediate states.  
Real particles scattering processes are also described through the quantum mechanism of
exchanges of virtual particles (gauge bosons, resonance states, more complex objects, such as Regge trajectories). In the diagram  technique, such particles are represented by the inner lines  in a diagram. 
At the same time, however,  the relationship between energy and momentum which holds for  real particles and is  expressed by the equation 
$p^2 = m^2c^4$  is no longer satisfied  in the intermediate states.
The concept of virtual off-mass-shell
particle is derivable from the microscopic violation of causality allowed
by the time-energy uncertainty relation \cite{WH, LP31, Heitler, BLP, HKN}.
The relation between the momentum and energy of a virtual particle may be anything required by the conservation of 4-momentum at the vertices.
In this regard, it should be noted that the content of the concept of ``virtual particle'' has undergone a significant change. Even in the recent past, virtual particles meant, as a rule, such particles in virtual states (for example, photons, electrons, pions), which were well studied in real states. A class of particles (quarks, gluons etc.) has appeared that, in principle, cannot be in real states due to the property of confinement in quantum chromodynamics. 
 
Although it is impossible to directly observe such intermediate states, their experimental study is of great interest and importance 
due to their nontrivial dynamic properties.
Nevertheless, a number of 
characteristics of the virtual particle can be measured indirectly.
The indirect measurement is a measurement in which the value of the unknown quantity sought is calculated using measurements of other quantities related to the measurand by some known relation \cite{M72}. 

The main goal of our  present research is twofold. 
First, to derive a formula for estimating the speed of a virtual particle. 
This is the subject of this Letter.
Second, based on the obtained formula, to develop and test a 
framework,
which allow, based on existing experimental data, to perform 
an  indirect measuring the  speed of virtual photons  in processes of
deep inelastic scattering  of electrons on protons \cite{FN3}.

The modern relativistic quantum theory in the description of particle interactions applies the Hamilton approach, where the momenta of particles play the dominant role. Moreover, at relativistic velocities, the speed of a particle is less informative than its momentum. And in the case of a real photon, knowing only its speed, nothing at all can be learned about its energy or momentum. 

A pertinent question arises: What is the value of data on the speed of a virtual particle?  Let us explain the essence of our idea with a few examples.

1) Let a body move in a gaseous or liquid medium. When the speed of the body is less than the speed of sound in the medium, the thermodynamic characteristics of the medium in front of the body change smoothly and continuously. However, if the speed of a body in a medium is close to or higher than that of sound, then a shock wave 
 appears in the form of a Mach cone, 
in the zone of which thermodynamic parameters
 of the medium (pressure, density, temperature, entropy) change abruptly, the additivity of wave processes is violated, and phase transitions can occur.

2) Let a charged particle moves uniformly through material media.
If the velocity of the relativistic particle exceeds the phase velocity of light in this medium, then the particle begins to emit electromagnetic radiation at a certain angle to the velocity vector. This is the Cherenkov-Vavilov effect \cite{Ch34, Ch37, W34}. Fermi calculations \cite{F40} for real media showed that Cherenkov-Vavilov radiation also causes an increase in ionization losses. For the processes of light emission by rapidly moving particles, the macroscopic properties of the medium are essential.

3) In 1904-1905 A. Sommerfeld in a series of articles \cite{Som1904, Som1905_1, Som1905_2} published the results  of studying the motion of an electron
 in the framework of Lorentz's theory of electromagnetism. It turned out that if the electron's speed $u$ is less than the speed of light in vacuum, $u <c$, then no external force is required to move with a constant speed. However, if 
$u> c$, then resistance to movement arises, and an external force is required for uniform movement. This resistance is due to the reaction to electromagnetic radiation that the electron emits.  However, it is not superfluous to note that most of these conclusions were made much earlier by O. Heaviside \cite{H1889, Tyap74}.

What these examples have in common is that if the particle speed exceeds a certain characteristic value, $ u_{cr} $, then  dynamics  of the process  changes dramatically. Thus, the magnitude of the particle velocity serves as a natural  indicator, thanks to which it becomes clear in which part of the phase space the process is taking place. 

The implementation of the ``uncertainty principle'' in Nature determines the existence of virtual particles with  a very wide dynamic range of characteristics. 
For this reason, a derivation of the formula for the speed of a virtual particle is based on the Heisenberg's uncertainty relations (HUR) and below, we first recall a number of basic facts and results of quantum mechanics. 

The uncertainty principle and uncertainty relations for observables of two canonically conjugate quantum mechanical operators found by Heisenberg  \cite{WH} are fundamental foundations of quantum mechanics.  And, as follows from J. von Neumann's theorem on hidden parameters 
Ref. \cite{JvN}, the removal of Heisenberg's uncertainty relations 
 from the theory destroys quantum mechanics.

In the article \cite{WH}, only a heuristic estimate was given of how the inaccuracy of  the particle coordinate, $ q_1 $,  is associated with the inaccuracy of the particle 
momentum, $ p_1 $, into one relation, $ p_1 q_1 \sim \hbar$, called the uncertainty relation. 
At the time, Heisenberg was a visiting scientist at Niels Bohr's institute in Copenhagen.
Three months later, Kennard \cite{FN0}  reformulated  \cite{Kennard27}  
Heisenberg's intuitive finding as a strict inequality relating the standard deviation of position and the standard deviation of momentum
and  showed how the right-hand side of HUR should be written.
H. Weyl provided another proof of HUR in Refs \cite{Weyl_g, Weyl_e}
and also gave the inequality a modern look,
\beq
\Delta p_x  \Delta x\geq \hbar/2.
\label{HUR}
\eeq

Soon, Robertson \cite{Rob} and then, 
in a more general form,  Schr${\rm\ddot{o}}$dinger \cite{Schrd} proved analogs of the 
inequality (\ref{HUR}) for the case of two arbitrary, not necessarily canonically 
conjugate, abstract operators. 
As a result, in most textbooks and books
on the basics of quantum mechanics, one of these proofs is presented  and 
the relation between canonically conjugate variables in the form (\ref{HUR})
is given  only for one of the projections (see for example Refs \cite{Schiff,  LLQM, Griff} and many others).

However, there are circumstances \cite{FN1}
 when one has to restrict oneself to estimating only  uncertainties 
of the \texttt{}modulus of  
certain vectors, say, the length of a three-dimensional domain, 
$|\mathbf{R}|$, and the particle momentum, $|\mathbf{P}|$, 
since there is no information about the vector projections. In this case, 
a different  relationship of uncertainties is needed, 
which has not been encountered in textbooks or scientific publications. 

In the following, we present 
in Sections 2 and 3 
a derivation of  new uncertainty relations  
and a formula for evaluating the  speed 
of  a virtual particle in indirect measurements. 
In 
Section 4 
we discuss how the metric characteristics of space and the speed of a virtual particle are interrelated in the  interaction region. 
An essence of our results is presented  in Summary.


\section{New  angular variable and uncertainty relation}
{\it New  angular variable and uncertainty relation.}---
Uncertainties of quantum mechanical
Hermitian operators $\hat{x}$ and $\hat{p_x}$ are defined 
(see Ref. \cite{Weyl_e}, p. 77 and  Ref. \cite{Griff}, p. 137) via
\beq
(\Delta x)^2={\mathrm \int^{+\infty}_{-\infty}}\!\! x^2{\bar{\varphi}}\varphi{ d}x,\,\,\,
(\Delta p_x)^2=\int^{+\infty}_{-\infty}\!\! {\bar{\varphi}}\frac{{\partial}^2\varphi}{{\partial}x^2}{ d}x.
\eeq
Therefore, we write out the uncertainty relations for all projections of the pair of conjugate coordinate-momentum variables in terms of mean square deviations,
\bea
(\Delta p_x)^2 (\Delta x)^2&\geq& (\hbar/2)^2, \nonumber \\
(\Delta p_y)^2 (\Delta y)^2&\geq& (\hbar/2)^2,\\
(\Delta p_z)^2 (\Delta z)^2&\geq& (\hbar/2)^2, \nonumber
\label{ve}
\eea
without extracting the square root, as in (\ref{HUR}).
If we now add the left-hand sides of these inequalities, 
we find that the resulting  sum is a scalar product 
$\mathbf{(\Delta P)^{(2)} \cdot (\Delta R)^{(2)}}$
of  the vectors 
$\mathbf{(\Delta P)^{(2)}}=((\Delta p_x)^2,(\Delta p_y)^2,(\Delta p_z)^2)$ 
and
$\mathbf{(\Delta R)^{(2)}}=((\Delta x)^2,(\Delta y)^2,(\Delta z)^2)$
from the Euclidean finite-dimensional sector of the Hilbert subspace, which includes
a set of vectors with only  positively defined projections.
However, the dot product introduces an angle between  vectors, and therefore for the 
norm of  vectors $\mathbf{(\Delta P)^{(2)}}$ and $\mathbf{(\Delta R)^{(2)}}$ 
we obtain the uncertainty relation, which includes the angular variable $\psi$,
\beq
\Vert \mathbf{(\Delta P)^{(2)}}\Vert\, \Vert \mathbf{(\Delta R)^{(2)}} \Vert\geq \frac{3\hbar^2}{4\cos \psi}.
\label{svp}
\eeq
The relations (3)  consist of 
only positive definite terms and this fixes the domain of the angle  $\psi\,\in[0, \pi/2)$, and 
the domain of the function values, $0<\cos {\psi} \le 1 $. 
Thus, depending on the state of the  physical system under study,  the value of $\cos{\psi}$
 varies and imposes restrictions on $\Vert \mathbf{(\Delta P)^{(2)}}\Vert$ and 
$\Vert \mathbf{(\Delta R)^{(2)}}\Vert$ of different degrees of stiffness.
%
The function $\cos{\psi}$ 
appears as a result of the reduction of six degrees of freedom in 
(3) to three degrees of freedom in (\ref{svp}) and reflects the geometric features of space (see Sec. 4).

Here are two specific examples of calculating $\cos{\psi}$.\\
1) 3D Harmonic Oscillator. With the use of the wave function of the 
n-th energy level, $\varphi_n$,  Ref. \cite{Griff},  vectors 
$\mathbf{(\Delta P)^{(2)}}$ and $\mathbf{(\Delta R)^{(2)}}$
are calculated exactly. For this problem we get $\cos{\psi}_{\,n} = 1/(2n+1)^2 $ with 
$n= 0,1,2,3,...$.
Thus, when $n\rightarrow \infty$, $\cos{\psi}_{\,n} \rightarrow 0$, and values of 
$\Vert \mathbf{(\Delta P)^{(2)}} \Vert$ and $\Vert \mathbf{(\Delta R)^{(2)}} \Vert$ become 
completely undefined.\\
2) 3D rectangular potential well with infinite walls, Ref. \cite{Griff}.
For this problem, calculations give $\cos{\psi}_{\,n} = 3/(n^2\pi^2-6) $, 
where $n= 1,2,3,...$. Again, in the limit $n\rightarrow \infty$, 
$\cos{\psi}_{\,n} \rightarrow 0$, and
values of 
$\Vert \mathbf{(\Delta P)^{(2)}} \Vert$ and $\Vert \mathbf{(\Delta R)^{(2)}} \Vert$ become completely undefined.\\
Note that for these physical systems the inequality sign in the relation 
(\ref{svp}) must be replaced with an equal sign.

 \section{Estimation of the particle speed from the uncertainty relations }
 
In the same 1927 article \cite{WH}, Heisenberg gives an uncertainty relation for 
another pair of canonically conjugate energy-time variables. This relation is definite only 
up to Planck's constant, so we write it out by including an arbitrary constant 
$\delta_H $,
\beq
(\Delta E)^2 (\Delta t)^2 \geq {\delta^2 \!\!\!}_{_H}\,\hbar^2 ,
\label{HB1}
\eeq
the value of which is fixed by the conditions of the problem being solved.
After lengthy and emotional discussions with Heisenberg
 these new discoveries in quantum mechanics  \cite{WHG, MK}, 
Niels Bohr presented  in a survey report in Como \cite{NB28}
 his own version of obtaining the  relation (\ref{HB1}). 
In this regard, the inequality (\ref{HB1}) is usually called the Heisenberg-Bohr relation.

Landau and Peierls  \cite{LP31}  generalized a number of conclusions of classical 
quantum mechanics to the relativistic domain. In particular, it was demonstrated that 
the Heisenberg inequalities for momentum and coordinate are also valid at relativistic 
velocities. In passing to the relativistic consideration, the inequality (\ref{HB1}), 
however, fails to give such a simple justification. Nevertheless, Landau and Peierls have 
derived  new inequalities for a free relativistic particle,
 Refs \cite{LP31}, \cite{LLQM}, \cite{FK47},
\beq
|\ur_i|\Delta p_i \Delta t \ge {\delta\!}_{_{LP}} \hbar,
\label{LP}
\eeq 
that holds for each of the components $i=(x,y,z)$ 
separately \cite{FN2}.
Here the symbol $\mathbf {u}$ denotes the group velocity vector of the particle, $\mathbf {u}=(u_x, u_y, u_z)$ and 
an arbitrary constant $\delta_{_{LP}} $ is introduced on the same 
reasoning as in the inequality (\ref{HB1}).
Adding the squares of the relations (\ref{LP}) for $ i = (x, y, z) $, 
like above, we obtain 
on the left-hand side of the inequality the scalar product 
$\mathbf{u^{(2)} \cdot (\Delta P)^{(2)}}$
of  the vectors $\mathbf{u^{(2)}}=((\ur_x)^2,(\ur_y)^2,(\ur_z)^2)$ and
$\mathbf{(\Delta P)^{(2)}}$.  In this way, we have identified a new angle  ${\psi}_{\,\ur}$
between the given vectors and get one more inequality connecting norms 
of  the particle  quadratic velocity vector, the mean square deviation of 
its momentum and the square of duration of the measurement process,
\beq
\Vert\mathbf{u^{(2)}}\Vert\,\Vert \mathbf{(\Delta P)^{(2)}}\Vert(\Delta t)^2
\geq 3(\delta_{_{LP}}\hbar)^2 /\cos {\psi}_{\,\ur}.
\label{LPL}
\eeq

The relations (\ref{svp}) and (\ref{LPL}) allows us to estimate the modulus of the particle group 
velocity  $|\mathbf{u}|$ under conditions when direct measurement of the velocity is impossible (the method of indirect measurements  \cite{M72}).
For this purpose, one need take the ratio  of the inequality (\ref{LPL})
to (\ref{svp}), and then replace $\Vert \mathbf {(\Delta R)^{(2)}}\Vert$ and 
$(\Delta t)^2$ according to Eqs (\ref{svp}) and (\ref{HB1}), respectively.
In this way,
\beq
\Vert \mathbf {u^{(2)}}\Vert\sim 4\frac{\Vert\mathbf{(\Delta R)^{(2)}}\Vert}{(\Delta t)^2}
\frac{\cos {\psi}}{\cos {\psi}_{\,\ur}}
=A\frac{(\Delta E)^2}{\Vert \mathbf{(\Delta P)^{(2)}}\Vert}.
\label{Vel}
\eeq
With the use of the Cauchy-Buniakowsky-Schwarz inequality we get
finally for the norm of speed the following estimate
\beq
\Vert \ur \Vert\le\sqrt{\sqrt{3}\Vert \mathbf{u^{(2)}}\Vert}=\sqrt{\sqrt{3}A\frac{(\Delta E)^2}{\Vert \mathbf{(\Delta P)^{(2)}}\Vert}}.
\label{Vel_f}
\eeq
Here $A= 3{\delta^2\!\!\!}_{_{LP}}/({\delta^2\!\!\!}_{_H}\, \cos {\psi}_{\,\ur})$. 

Before clarifying the connection between the speed of a virtual particle, the magnitude of the normalization constant, $A$ and the metric properties of space, we list below a number of  
facts from the theory of metric spaces. 

\section{Space-time metrics in the interaction region }

Minkowski space plays an important role in high energy physics 
and is a special case of Euclidean space with an indefinite metric (IM), also called 
pseudo-Euclidean space. 
Moreover,  Minkowski space is the simplest example of a pseudo-Riemannian space. The pseudo-Riemannian geometry (built like the Riemannian one, but with IM) underlies the mathematical apparatus of the general theory of relativity. 
$n$-dimensional Euclidean, pseudo-Euclidean, Riemannian, and pseudo-Riemannian spaces are special cases of Hilbert spaces with definite or indefinite metrics \cite{Kurosh88, Rash67}.  Hilbert space serves as the mathematical basis of the apparatus of quantum mechanics and  quantum field theory.

To classify (pseudo-)Euclidean spaces, the so-called space index $ k $ (or the index of inertia) is 
introduced, defined as the number of imaginary unit basis vectors of the 
orthonormal frame \cite{Rash67, Shafarevich13}. For the proper Euclidean space, the space index 
is $ k = 0 $. For  Minkowski space with total dimension $n = 4$ and signature 
(+,-,-,-)=(1, 3), the space index is $k=3$.
As applied to  problems of physics, in Minkowski space one coordinate is interpreted as time, and  other coordinates correspond to 3D Euclidean space. 

Let us now discuss the condition for the normalization of the constant $A$ in  formula 
(\ref{Vel_f}). In a subsequent article \cite{FN3}, we are going to  apply  formula (\ref{Vel_f}) to estimate the speed of virtual photons from data on deep inelastic $ ep $ scattering at the HERA collider.
The magnitude of the photon virtuality, $ Q^2 $,  varies in a very wide range, and in the limit 
$ Q^2 \rightarrow 0$ GeV$^2$ (a real photon limit) there should be 
$\Vert \ur \Vert \rightarrow c$. This  condition allows to fix  the value of the constant $ A $. 
 
In the Standard Model, it is assumed that the geometry of space-time  known to us from macroscopic physics remains applicable in the microcosm too.
The derivation of the inequality (\ref{svp}) was based on this assumption, where the formula for the scalar product of vectors in the proper Euclidean space was applied, 
$$(\mathbf{ab})=|a||b|\cos \psi .$$
The inequality (\ref{LP}) comes from the  inequality (\ref{HB1}). Therefore,
there is  a good reason to believe that ${\delta^2\!\!\!}_{_{LP}}={\delta^2\!\!\!}_{_H}$. In this case  $ A = 3 / \cos {\psi} _ {\, \ur} $. 
Consequently, if it follows from an experimental data that $ A \ge 3 $,
then 3D Euclidean geometry is preserved in the interaction region 
and the space index  is equal to $ k = 3 $. 

The case is quite different if $ A <3 $. Then our assumption about the metric of the Euclidean part of  Minkowski space is incorrect. 
The three-dimensional subspace, where  vectors  $\mathbf{a}=\mathbf{(\Delta P)^{(2)}}$
and $\mathbf{b}=\mathbf{(\Delta R)^{(2)}}$
are defined, has a pseudo-Euclidean (or pseudo-Riemannian) metric with a scalar product
$$(\mathbf{ab})=|a||b|\cosh \psi$$ (see \cite{Rash67}, Sec. 46).
Therefore, $A= 3/ \cosh {\psi}_{\,\ur}$ and the space index is  $ k = 2 $.

In  nonlinear field theories (with scalar or/and vector fields) the signal velocity depends on 
the magnitude of the field and its derivatives  \cite{DBlo52, DBloOrl53, DBlo57, DBlo73}. 
In this case, the speed of the signals can be less or more 
than the speed of light $ c $ in void and the space index change from $k=3$ to $k=2$ or even $k=1$  \cite{DBloOrl53, DBlo66, DBlo73, DBlo74}. 
This corresponds to the space-time domain, in which there are three time dimensions:
$t_1=x_1$, $t_2=x_2$, $t_3=x_3$  and one spatial dimension $x=x_4$.

As applied to our problem, this  mean that in the interaction region, the metric characteristics of space differs from the Minkowski metric  due to the non-linearity of  processes 
and  high intensity of  fields.

\section{SUMMARY }

We propose a new method for revealing the type of space-time geometry 
in the interaction region of high-energy particles. The method is  based on
new uncertainty  relations (\ref{svp}) and (\ref{LPL}), created 
by reducing the number of degrees of freedom in  uncertainty relations (URs) 
obtained by Heisenberg and  by Landau and Peierls. 
In this form,  URs directly demonstrate 
a connection with the metric of space in the region where the particle is located. 
On the basis of the relations obtained, we derive a formula for estimating 
the speed  of a virtual particle in indirect measurements.
The normalization constant $A$ contains
 information about the space metric, 
and its value is fixed in the limit of zero virtuality of the particle. 
 If  $ A \ge 3 $, then 3D Euclidean geometry is preserved in the interaction region and 
the space index  is $ k = 3 $.
However,  if $ A <3 $ then the three-dimensional subspace has a pseudo-Euclidean (or even pseudo-Riemannian) metric with the space index   $ k = 2 $.
It should be noted, however, that within the framework of our approach, we cannot answer the question: whether there is a region with a changed geometry around the particle initially or whether the change in the metric occur  during the  interaction.

To conclude,  let us give one example how works the method of indirect measurements.
For instance,  based on the characteristics of final state particles and groups of particles (jets)  
from experiments 
at the LHC, HERA etc. colliders, the virtual part of  dynamics of strong and electroweak interactions is experimentally studied and the momentum fraction distributions of virtual quarks and gluons are reconstructed via measurements  of  structure functions \cite{DESY_15_039}. 

\section*{ACKNOWLEDGMENTS}

The author is grateful to M. Wing for reading an early version of  the manuscript and
suggestions, E. Boos, A. Geiser, N. Nikitin, E. Oborneva,  I. Volobuev  
and other colleagues from SINP 
and the ZEUS collaboration for discussions and comments.
The author is especially grateful to K. G. Gulamov and N. N. Nikolaev for their mentorship in the early days of this  amazing scientific journey.

%
\bibliography{hiddenen}

\end{document}